\documentclass[a4paper,12pt]{article}

\makeatletter

\@addtoreset{equation}{section}
\makeatother

\usepackage{epsfig}
\usepackage{graphicx}
\usepackage{amsmath}
\usepackage{amssymb}

\begin{document}
\begin{titlepage}
\null
\begin{flushright}
TIT/HEP-599  \\
December, 2009
\end{flushright}

\vskip 1.8cm
\begin{center}

 {\Large \bf On Non-linear Action for Gauged M2-brane}

\vskip 2.3cm
\normalsize
\renewcommand\thefootnote{\alph{footnote}} 

  {\bf Shin Sasaki\footnote{shin-s(at)th.phys.titech.ac.jp}}

\vskip 0.5cm

{\normalsize\it Department of Physics\\
Tokyo Institute of Technology\\
Tokyo, 152-8551, Japan} 

\vskip 3cm

\begin{abstract}
We propose a non-linear extension of $U(1) \times U(1)$ (abelian) 
ABJM model 
including $T_{M2}$ (higher derivative) corrections. 
The action proposed here is 
expected to describe a single M2-brane proving $\mathbf{C}^4/\mathbf{Z}_k$ 
target space. 
The model includes couplings with the 3-form background
in the eleven-dimensional supergravity which is consistent with the 
orbifold projection. 
We show that the novel higgs mechanism proposed by Mukhi and 
Papageorgakis does work even in the presence of higher derivative 
corrections and couplings with the background field, giving the correct structure 
of the Dirac-Born-Infeld action with Wess-Zumino term for a D2-brane.
We also find half BPS solutions in the full non-linear theory which is 
interpreted as an another M2-brane intersecting with the original 
M2-brane. A possible generalization to $U(N) \times U(N)$ gauge group is briefly discussed.
\end{abstract}

\end{center}
\end{titlepage}

\newpage
\setcounter{footnote}{0}
\renewcommand\thefootnote{\arabic{footnote}} 
\section{Introduction}
Recently, the low-energy description of multiple M2-branes has been 
intensively studied. 
Bagger, Lambert and Gustavsson proposed a three dimensional $\mathcal{N} 
= 8$ supersymmetric model with $SO(8)_R$ R-symmetry (BLG model) as an effective 
theory of multiple M2-branes which is based on the three-algebra 
structure \cite{BaLa, Gu}. 
On the other hand, Aharony-Bergman-Jafferis-Maldacena recently proposed 
a three dimensional $\mathcal{N} = 6$ superconformal Chern-Simons-matter theory as an 
alternative model of $N$ coincident 
M2-branes probing $\mathbf{C}^4/\mathbf{Z}_k$ orbifold (ABJM model) \cite{AhBeJaMa}. 
The gauge group is $U(N) \times U(N)$ and four complex scalar fields are 
introduced as the (anti) bi-fundamental representation of 
the gauge group. Two gauge fields $A_{m}$ and $\hat{A}_{m}$ 
corresponding to each $U(N)$s have levels $k$ and $-k$ respectively and 
the Lagrangian of the ABJM model exhibits an $SU(4)_R$ R-symmetry. 
For $SU(2) \times SU(2)$ gauge group, the $\mathcal{N} = 6$ 
supersymmetry is enhanced to $\mathcal{N} = 8$ and the ABJM model 
coincides with the BLG model \cite{GuSJ}. It is also argued in \cite{AhBeJaMa} that 
the model is dual to M-theory on $AdS_4 \times S^7/\mathbf{Z}_k$ at large-$N$.

A lot of works on the classical solutions of this model have been achieved.
In \cite{Te}, a BPS fuzzy funnel configuration that represents an 
M5-brane intersecting with multiple M2-branes was found\footnote{See 
\cite{KrMa} for 
the related work in the BLG model.}. 
A domain wall solution that interpolates between a fuzzy sphere and trivial vacua was found 
in the mass deformed ABJM model \cite{HaLi}. 
There is a detail discussion on the dimensionality of the fuzzy 
sphere \cite{NaPaRa}. Other BPS objects such as vortices, Q-balls and so on were studied in 
\cite{ArMoSa, KaSa, KiKiKwNa, AuKu}.

It is known that M2-branes in eleven dimensions are reduced to D2-branes 
in ten dimensions by compactifing one of the transverse direction to the 
M2-branes. This procedure is performed by the novel higgs mechanism proposed in \cite{MuPa}.
It has been shown that the $U(N) \times U(N)$ ABJM theory is reduced to 
$U(N)$ super Yang-Mills theory describing $N$ coincident D2-branes. Since the super Yang-Mills 
action is the leading order approximation in $\alpha'$ (string scale) expansion of 
the non-abelian Dirac-Born-Infeld (DBI) action, 
it is natural to expect that the ABJM action is modified by the higher 
derivative membrane corrections that are suppressed by the 
eleven-dimensional Planck mass $M_{11}$ or M2-brane tension $T_{M2}$.
Various investigations on this issue have been carried out especially in 
the (Lorentzian) BLG models \cite{AlMu, IeRu, EzMuPa, Ga}, while 
corresponding studies in the ABJM model have not been analyzed so far.

In this paper we investigate higher derivative corrections in the ABJM model, 
namely, $T_{M2}^{-1}$ corrections in eleven dimensions. 
Since it is known that the non-abelian DBI action \cite{My, Ts} 
is not correct at least at $\mathcal{O} (F^6)$ \cite{Ba}, we 
focus on the abelian DBI action which is correct for all orders in $\alpha'$. 
This abelian DBI action should be obtained from $U(1) 
\times U(1)$ (abelian) ABJM model including all order of $T_{M2}^{-1}$ corrections. 
We find a simple non-linear generalization of the abelian ABJM model that correctly 
reproduces the abelian DBI action after compactifing one transverse 
direction. In this process, the non-dynamical gauge fields $A_m, 
\hat{A}_m$ become dynamical via the novel higgs mechanism. 
We also consider a coupling of a single M2-brane with the 3-form in the 
eleven-dimensional supergravity. This coupling is given by the 
Wess-Zumino term which is covariantized by the 
world-volume gauge group. The Wess-Zumino term reproduces the 
correct structure of the couplings with the NS-NS 2-form and R-R 3-form 
backgrounds in the D2-brane DBI action.

To confirm the availability of this non-linear action, we derive 1/2 
BPS conditions both in the abelian ABJM and its non-linear extension. 
We demonstrate that the BPS solution in the ABJM model is an exact solution 
even in the non-linear level. All the higher derivative effects cancel 
out provided that the BPS configuration is concerned. This result is 
consistent with the D-brane cases in string theories.

The organization of this paper is as follows. 
In the next section, we briefly explain the D2-brane effective theory.
In section 3, the abelian ABJM model in eleven dimensions 
and its reduction to the D2-brane action in type 
IIA string theory are reviewed. A BPS solution preserving half 
supersymmetry is derived in this section. The solution is interpreted as 
an M2-brane intersecting with the original M2-brane. 
In section 4, we introduce a non-linear generalization of the abelian 
ABJM model that couples to the 3-form background in the eleven-dimensional supergravity.
We show that by the higgs mechanism, the model 
is adequately reduced to the abelian DBI action with Wess-Zumino term. 
Section 5 is conclusions and discussions where non-abelian generalization 
is briefly discussed.

\section{D2-brane effective theory}
In this section, we review the effective action of a single D2-brane in 
the DBI and its dual forms. If the derivative corrections to the gauge field 
strength $F_{mn}$ is ignored, the abelian DBI action contains all 
$\alpha'$ corrections leading to the resolution of the field strength 
singularity \cite{BoIn} and allows supersymmetric extension of the 
action \cite{Ts}.
The bosonic part of the single D2-brane effective action is given by the 
following DBI form,
\begin{eqnarray}
S_{D2} = - T_{D2} \int \! d^3 x \ \sqrt{- \det (P[g]_{mn} + \lambda F_{mn} )},
\label{D2DBI}
\end{eqnarray}
where $P[g]_{mn}$ is the induced metric and $F_{mn} = \partial_m A_n - 
\partial_n A_m, \ m,n = 0, 1, 2$ is the field strength of the 
world-volume $U(1)$ gauge field $A_m$.
$\lambda = 2\pi \alpha'$ and $T_{D2} = \frac{1}{(2\pi \alpha')^{2}} 
g^{-1}_{\mathrm{st}} \alpha^{\prime - \frac{3}{2}}$ is the tension of a D2-brane. 
$g_s$ is the string coupling constant. 
In this paper we consider flat Minkowski background and neglect 
 curvature couplings of gravity. 
In the static gauge, the pull-back of the metric is 
evaluated as 
\begin{eqnarray}
P[g]_{mn} = \eta_{mn} + \lambda^2 \sum_{I=1}^7 \partial_m X^I \partial_n X^I
= \eta_{mn} + \frac{T_{D2}^{-1}}{g^2_{D2}} \sum_{I=1}^7 \partial_m X^I \partial_n X^I,
\end{eqnarray}
where $\eta_{mn} = (-1,+1,+1)$ and $X^I, \ (I = 1, \cdots 7)$ are scalar fields representing fluctuations in the 
transverse directions to the D2-brane. 
Here we have used the relation that the gauge coupling constant $g_{D2}$ 
in the D2-brane world-volume theory is defined by $T_{D2} \lambda^2 = \frac{1}{g^2_{D2}}$.
The action (\ref{D2DBI}) preserves $\mathcal{N} = 8$ supersymmetry (16 
supercharges) and is invariant 
under the $SO(7)_R$ R-symmetry and $U(1)$ gauge symmetry. 
Since the action (\ref{D2DBI}) is highly non-linear, it is useful to 
linearize it by introducing a Lagrange multiplier $p$. 
The action (\ref{D2DBI}) is rewritten as 
\begin{eqnarray}
S_{D2} = \int \! d^3 x \ \left[
\frac{p}{2} \det g - \frac{T^2_{D2}}{2p} \det 
(\delta_m {}^n - \lambda F_{mp} g^{pn} )
\right],
\label{D2DBI2}
\end{eqnarray}
where we have omitted the symbol ``$P$'' of the pull-back.
It is easy to check that the original action (\ref{D2DBI}) is 
recovered by solving the equation of motion for $p$
and put the solution back into the action (\ref{D2DBI2}). 
The action (\ref{D2DBI2}) can be further rewritten by 
the introduction of an auxiliary field $t_m$, 
\begin{eqnarray}
S_{D2} = \int \! d^3 x \ \left[
\frac{p}{2} 
\det g \left(1 + \lambda^2 g^4_{D2} t_m t_n g^{mn} \right) 
+ \frac{1}{2} \epsilon^{mnp} t_m F_{np} - \frac{T^2_{D2}}{2p}
\right].
\label{D2DBI3}
\end{eqnarray}
Again, one can show that the action (\ref{D2DBI2}) is recovered once the 
auxiliary field is integrated out.
At the leading order in $\alpha'$, the action (\ref{D2DBI})
reduces to that of the (bosonic part of) the supersymmetric Maxwell action.

On the other hand, once the D2-brane couples to the NS-NS 2-form $B$ and 
R-R 3-form $C^{(3)}$ backgrounds in type IIA supergravity, the effective action is 
given by 
\begin{eqnarray}
S_{D2} = - T_{D2} \int \! d^3 x \ \sqrt{- \det (P[g]_{mn} + P[B]_{mn} + \lambda 
F_{mn} )} + T_{D2} \int \! d^3 x \ \epsilon^{mnp} P[C^{(3)}]_{mnp},
\label{D2DBIWZ}
\end{eqnarray}
where the pull-back of $B_{mn}$ in the static gauge is evaluated as 
\begin{eqnarray}
P[B]_{mn} = B_{mn} + \lambda B_{mI} \partial_n X^I 
- \lambda B_{nI} \partial_m X^I + \lambda^2 B_{IJ} \partial_m X^I 
\partial_n X^J.
\end{eqnarray}
Here $I,J = 1, \cdots, 7$ label the transverse directions to the 
D2-brane world-volume and $\epsilon^{ij}, \epsilon_{ij}$ are 
anti-symmetric tensors with $\epsilon^{12} = - \epsilon_{12} = 1$.
The pull-back of the R-R 3-form is evaluated as 
\begin{eqnarray}
P[C^{(3)}]_{mnp} = 
C^{(3)}_{mnp} + 3 \lambda C^{(3)}_{mnI} \partial_p X^I 
+ 3 \lambda^2 C^{(3)}_{m IJ} \partial_n X^I \partial_p X^J
+ \lambda^3 C^{(3)}_{IJK} \partial_m X^I \partial_n X^J \partial_p X^K.
\label{RR3}
\end{eqnarray}
At leading order in $\alpha'$, interaction terms in the DBI Lagrangian $\mathcal{L}_{DBI}$
in (\ref{D2DBIWZ}) which contain at least one NS-NS 2-form are 
\begin{eqnarray}
\mathcal{L}_{DBI} &\sim& 
- \frac{T_{D2}}{4} B_{mn} B^{mn} - \lambda T_{D2} B^{mn} 
B_{mI} \partial_n X^I - \frac{T_{D2} \lambda}{2} F^{mn} B_{mn}
- \frac{1}{2 g^2_{D2}} B^{mn} B_{IJ} \partial_m X^I \partial_n X^J 
 \nonumber \\
& & 
- \frac{1}{2 g^2_{D2}} B_{mI} B^m {}_J \partial_n X^I \partial^n X^J 
+ \frac{1}{2 g^2_{D2}} B_{mI} B_{nJ} \partial^n X^I \partial^m X^J 
- \frac{1}{g^2_{D2}} F^{mn} B_{mI} \partial_n X^I
\nonumber \\
& & + \mathcal{O} (T_{D2} \lambda^3).
\label{NSNS2}
\end{eqnarray}
We will see that the first line in the above expression is projected out 
in eleven dimensions 
and the second line will be obtained from the ABJM model accompanied with the 
coupling with 3-form in eleven-dimensional supergravity.

\section{$U(1) \times U(1)$ ABJM model}
\subsection{Reduction to ten dimensions via the higgs mechanism}
If one of the transverse direction to an M2-brane in eleven-dimensional 
M-theory is compactified, the M2-brane reduces to a D2-brane in 
ten-dimensional type IIA string theory. This procedure from the 
viewpoint of the M2-brane effective theory was proposed, first in the BLG model 
\cite{MuPa} and later performed in the ABJM model \cite{PaWa}. 
Assuming that one of the four scalar fields in the ABJM model develops 
a real VEV\footnote{Note that this VEV should be a flat direction of vacua.} $v$ and taking 
the limits $v, k \to \infty$ with fixed $v/k$, the cone in the target 
space is regarded as a cylinder at points far from the origin and 
which substantially leads to the compactification of the transverse direction. 
The non-dynamical gauge fields become the dynamical one in this 
procedure and the $\mathcal{N} = 8$ supersymmetric $U(N)$ gauge theory 
is obtained from the $U(N) \times U(N)$ ABJM model. This mechanism is 
known as the novel higgs 
mechanism. In the rest of this subsection, we review this mechanism by 
focusing on the abelian ABJM model. 
The $U(1) \times U(1)$ ABJM action \cite{AhBeJaMa, BeKlKlSm} is given by 
\begin{eqnarray}
S_{\mathrm{ABJM}} &=& \int \! d^3 x \ \left[
\frac{k}{4\pi} \epsilon^{m n p} (A_{m} 
\partial_{n} A_{p} - \hat{A}_{m} \partial_{n} \hat{A}_{p}) 
- D_{m} Y^{\dagger}_A D^{m} Y^A - i \psi^{\dagger A} \gamma^m D_m \psi_A
\right],
\label{ABJM_action}
\end{eqnarray}
where an integer $k$ is the Chern-Simons level, $\gamma^m$ are the 
three-dimensional gamma matrices, 
$Y^A \ (A=1, \cdots, 4)$ are complex scalars transforming in 
$\mathbf{4}$ by the $SU(4)_R$ R-symmetry and $\psi_A$ are superpartner 
of $Y^A$. The gauge covariant derivative is defined as 
$D_{m} Y^A = \partial_{m} Y^A + i A_{m} Y^A - i Y^A \hat{A}_{m}$. 
The action is gauge invariant under the following $U(1) \times U(1)$ gauge 
transformation, 
\begin{eqnarray}
Y^A \to e^{i \lambda} Y^A e^{- i \hat{\lambda}}, \ 
\psi_A \to e^{i \lambda} \psi_A e^{- i \hat{\lambda}}, \ 
\ A_{m} 
\to A_{m} - i \partial_m \lambda, \ \hat{A}_m \to \hat{A}_m - i \partial_m 
\hat{\lambda}, 
\label{gauge_transf} 
\end{eqnarray}
where $\lambda, \hat{\lambda}$ are gauge parameters for 
each $U(1)$ group. This action represents a single M2-brane probing $\mathbf{C}^4/\mathbf{Z}_k$ 
orbifold and has $\mathcal{N}=6$ supersymmetry in three dimensions. 
The orbifold projection is defined by $Y^A \to e^{\frac{2\pi i}{k}} Y^A$.
The $\mathcal{N} = 6$ supersymmetry transformation is given by \cite{Te}
\begin{eqnarray}
\begin{aligned}
 & \delta Y^A = i \omega^{AB} \psi_B, \\
 & \delta Y^{\dagger}_A = i \psi^{\dagger B} \omega_{AB}, \\
 & \delta \psi_A = - \gamma^m \omega_{AB} D_m Y^B, \\
 & \delta \psi^{\dagger A} = D_m Y^{\dagger}_B \omega^{AB} \gamma^m, \\
 & \delta A_m = \frac{\pi}{k} 
\left(
- Y^A \psi^{\dagger B} \gamma_m \omega_{AB} + \omega^{AB} \gamma_m 
\psi_A Y^{\dagger}_{B}
\right), \\
 & \delta \hat{A}_m = \frac{\pi}{k} 
\left(
- \psi^{\dagger A} Y^B \gamma_m \omega_{AB} + \omega^{AB} \gamma_m 
Y^{\dagger}_A \psi_B
\right),
\end{aligned}
\end{eqnarray}
where $\omega^{AB}, \omega_{AB}$ are supersymmetry parameters. 
In the following, we consider only the bosonic part of the action and keep 
$k \gg 1$ so that the classical analysis is reliable. 
Let us see that this action reduces to the $\mathcal{N} = 8$ 
supersymmetric $U(1)$ gauge theory by the higgs mechanism.
We consider the following decomposition, 
\begin{eqnarray}
A'_m = \frac{1}{2} (A_m + \hat{A}_m), \qquad B'_m = \frac{1}{2} (A_m 
- \hat{A}_m).
\end{eqnarray}
Once a scalar field, for example $Y^4$, develops a VEV $v \in \mathbf{R}$,
the gauge symmetry is broken down to its diagonal part, $U(1) \times U(1) \to 
U(1)_{\mathrm{diag}}$ and the scalars become gauge neutral. After shifting the scalar field
$Y^A \to v \delta^A {}_4 + Y^A $ and rescaling $B'_{m} \to B'_{m}/v$ to 
keep the finite kinetic term of the gauge field, 
and decomposing the complex scalar fields as 
\begin{eqnarray}
Y^A = \frac{1}{\sqrt{2}} (X^A + i X^{A+4}), \quad (A = 1, \cdots, 4),
\end{eqnarray}
where $X^I \ (I=1, \cdots, 8)$ are real, the action becomes 
\begin{eqnarray}
S_{\mathrm{ABJM}} &=& \int \! d^3 x \ \left[
\frac{k}{2\pi v} \epsilon^{m n p} F'_{m 
n} B'_{p} -  \partial_{m} Y^{\dagger}_A \partial^{m} Y^A 
-  \frac{2i}{v} B'_{m} Y^A \partial^{m} Y^{\dagger}_A + \frac{2i}{v} 
B'_{m} Y^{\dagger}_A \partial^{m} Y^A 
\right.
\nonumber \\
& & \left. \qquad 
- \frac{4}{v^2} B'_{m} B^{\prime m} Y^A Y^{\dagger}_A 
-  \frac{4}{\sqrt{2}} B'_{m} \partial^{m} X^8 - 
 \frac{8i}{\sqrt{2}v^2} B'_{m} B^{\prime m} 
X^4 - 4 B'_{m} B^{\prime m}
\right],
\end{eqnarray}
where $F'_{m n} = \partial_{m} A'_{n} - \partial_{n} A'_{m}$.
Then taking the limit $v, k \to \infty$ with $k/v$ fixed, a transverse 
direction is compactified and the effective action reduces to that of 
a D2-brane in ten-dimensional type IIA string theory. The resulting 
action $S_{D2}$ is 
\begin{eqnarray}
S_{D2} = \int \! d^3 x \ \left[
\frac{k}{2\pi v} \epsilon^{m n p} F'_{m n} 
B'_{p} - \partial_{m} Y^{\dagger}_A \partial^{m} Y^A -  
 \frac{4}{\sqrt{2}} B'_{m} \partial^{m} X^8
- 4 B'_{m} B^{\prime m}
\right].
\end{eqnarray}
Since $B'_{m}$ is the auxiliary field, it can be integrated out giving 
the bosonic part of the $\mathcal{N} = 8$ supersymmetric $U(1)$ gauge theory with manifest $SO(7)_R$ R-symmetry,
\begin{eqnarray}
S_{D2} = \int \! d^3 x \ \frac{1}{g^2_{D2}} \left[
- \sum_{I = 1}^7 \frac{1}{2} \partial_{m} X^I \partial^{m} X^I - 
\frac{1}{4} F'_{m n} F^{\prime m n}
\right],
\end{eqnarray}
where at the final step, we have rescaled $X^I \to g_{D2}^{-1} X^I$ 
and defined the gauge coupling constant in D2-brane as 
$\frac{1}{g^2_{D2}} \equiv \frac{k^2}{8 \pi^2 v^2}$. 

\subsection{BPS solutions}
Since BPS configurations in effective theories of branes frequently 
become useful guidelines toward the construction of a non-linear action, 
we study BPS solutions of the abelian ABJM model. 
One can show that the energy $E$ of the abelian ABJM model is given by 
\begin{eqnarray}
E &=& \int \! d^2 x \ \left[ |D_0 Y^A|^2 + |D_i Y^A|^2 \right]
\nonumber \\
&=& \int \! d^2 x \
\left[
|D_0 Y^A|^2 + \frac{1}{2} 
\left|
D_i Y^A \pm i \epsilon_{ij} D^j Y^A
\right|^2
\right]
\pm i \int \! d^2 x \ \epsilon^{ij} D_i Y^A D_j Y^{\dagger}_A,
\end{eqnarray}
where $i,j$ are the space indices in the M2-brane world-volume. 
The last term is rewritten as 
\begin{eqnarray}
\pm i \int \! d^2 x \ \epsilon^{ij} D_i Y^A D_j Y^{\dagger}_A 
= \pm i \epsilon^{ij} \int \! d^2 x \ 
\partial_i (Y^A D_j Y^{\dagger}_A) 
\mp 
\frac{i}{2} \epsilon^{ij} \int \! d^2 x \ 
i |Y^A|^2 (\hat{F}_{ij} - F_{ij}).
\end{eqnarray}
Since the gauge fields are non-dynamical, 
constraints come from the equations of motion for the gauge fields. 
These are given by 
\begin{eqnarray}
\begin{aligned}
 & \frac{k}{4\pi} \epsilon^{mpq} F_{pq} =  i (Y^A D^m Y^{\dagger}_A - 
Y^{\dagger}_A D^m Y^A),  \\
 & \frac{k}{4\pi} \epsilon^{mpq} \hat{F}_{pq} =  i (Y^A D^m Y^{\dagger}_A - 
Y^{\dagger}_A D^m Y^A).
\end{aligned}
\label{CS_constraint}
\end{eqnarray}
We call these Chern-Simons constraints. 
From the constraints, we have the relation $F_{12} - \hat{F}_{12} = 0$. 
Then the energy bound is obtained as 
\begin{eqnarray}
E &=& \int \! d^2 x \ 
\left[
|D_0 Y^A|^2 + \frac{1}{2} 
\left|
D_i Y^A \pm i \epsilon_{ij} D^j Y^A
\right|^2
\right]
\pm i \epsilon^{ij} \int \! d^2 x \ \partial_i (Y^A D_j Y^{\dagger}_A) 
\nonumber \\
&\ge& \pm i \epsilon^{ij} \oint \! d x_i \ Y^A D_j Y^{\dagger}_A
\end{eqnarray}
The equality holds if the following BPS equations are satisfied,
\begin{eqnarray}
\begin{aligned}
 & D_0 Y^A = 0, \\
 & D_1 Y^A \pm i D_2 Y^A = 0.
\label{BPSeq}
\end{aligned}
\end{eqnarray}
The Chern-Simons constraints are rewritten, in terms of $A'_m, B'_m$, as 
\begin{eqnarray}
\begin{aligned}
 & \frac{k}{2\pi} \epsilon^{mpq} F^{(B)}_{pq} = 0, \\
 & \frac{k}{2\pi} \epsilon^{mpq} F'_{pq} =  2 i (Y^A D^m Y^{\dagger}_A 
- Y^{\dagger}_A D^m Y^A),
\end{aligned}
\end{eqnarray}
where $F^{(B)}_{pq} = \frac{1}{2} (F_{pq} - \hat{F}_{pq}), 
F'_{pq} = \frac{1}{2} (F_{pq} + \hat{F}_{pq})$. 
Assuming the configuration $Y^1 = Y \not=0, Y^A = 0 \ (A = 2,3,4)$, and 
taking the gauge $B'_m = 0$, the BPS equations (\ref{BPSeq}) reduce to 
\begin{eqnarray}
\begin{aligned}
 & \partial_0 Y = 0, \\
 & \bar{\partial} Y = 0, \quad \mathrm{or} \quad \partial Y = 0,
\end{aligned}
\end{eqnarray}
where we have defined $z = \frac{1}{\sqrt{2}} (x_1 + i x_2)$ and 
$\partial = \frac{\partial}{\partial z}$.
From the first condition, we find that $Y$ is time independent.
Therefore $Y$ is a static (anti)holomorphic function. 
This is just the solution 
discussed in \cite{GaGoTo, CaMa} where the solution is interpreted as an 
M2-brane intersecting with an another M2-brane. 
In \cite{GaGoTo, CaMa}, holomorphic embeddings of M2-branes without 
gauge fields were analyzed while here, in addition to the scalar field 
$Y$, we have the non-trivial gauge field configuration. We will see 
that this gauge field configuration is consistent with the M2-brane 
interpretation of the solution. 

Let us analyze the gauge field sector of the solution. 
Once we employ the gauge $B'_m = 0$ the Chern-Simons constraints become
\begin{eqnarray}
\frac{k}{4\pi} \epsilon^{mpq} F'_{pq} =  i (Y \partial^m Y^{\dagger} - 
Y^{\dagger} \partial^m Y).
\label{CS_constraint_gauge}
\end{eqnarray}
Because $Y$ is time independent, we have the following result,
\begin{eqnarray}
\epsilon^{0ij} F'_{ij} = 0.
\end{eqnarray}
Accordingly, the magnetic field $\tilde{B}$ is zero. Here we 
have defined the electric field $E'_i$ and the magnetic field 
$\tilde{B}$ as 
\begin{eqnarray}
F'_{pq} = 
\left(
\begin{array}{ccc}
0 & E'_1 & E'_2 \\
- E'_1 & 0 & \tilde{B} \\
- E'_2 & - \tilde{B} & 0
\end{array}
\right).
\end{eqnarray}
From the space components in the Chern-Simons constraint (\ref{CS_constraint_gauge}), we have 
\begin{eqnarray}
E_{z} =  \frac{2\pi}{k} (Y \bar{\partial} Y^{\dagger} - Y^{\dagger} 
\bar{\partial} Y),
\end{eqnarray}
where we have defined
\begin{eqnarray}
E_z = \frac{1}{\sqrt{2}} (E'_1 + i E'_2) = \partial_0 A'_z - 
\bar{\partial} A'_0, \qquad A'_z = \frac{1}{\sqrt{2}} (A'_1 + i A'_2).
\end{eqnarray}
Because $Y$ is time independent, $E_z$ should be too. Therefore we have $\partial_0 A'_z = 0$
\footnote{The most general solution for $A'_z$ is $\partial_0 A'_z = f (z, 
\bar{z})$ where $f$ is a time independent function. Here we assume $f (z, 
\bar{z}) = 0$ for simplicity.}, and then 
$A_0'$ is time independent. Once we choose $Y$ as a holomorphic function, 
the Chern-Simons constraint implies  
\begin{eqnarray}
\bar{\partial} A_0' = - \frac{2\pi}{k} Y \bar{\partial} Y^{\dagger}.
\end{eqnarray}
A solution to this constraint is given by
\begin{eqnarray}
A'_0 = - \frac{2\pi}{k} |Y|^2 + \mathrm{const}.
\end{eqnarray}
On the other hand, when $Y$ is an anti holomorphic function, the 
Chern-Simons constraint is  
\begin{eqnarray}
\bar{\partial} A_0' = + \frac{2\pi}{k} Y^{\dagger} \bar{\partial} Y.
\end{eqnarray}
A solution is 
\begin{eqnarray}
A'_0 = + \frac{2\pi}{k} |Y|^2 + \mathrm{const}.
\end{eqnarray}
Since we have the condition $\tilde{B} = 0$, $A'_z$ should be in  the pure gauge. 
In the D2 limit, $A'_m = \frac{1}{2} (A_m + \hat{A}_m) $ becomes 
dynamical and $B'_m = \frac{1}{2} (A_m - \hat{A}_m)$ is decoupled. 
As discussed in \cite{TeYa}\footnote{In \cite{TeYa}, a non-BPS D2-F1-D0 bound state was 
discussed in the abelian ABJM model.}, in the D2-limit, $A'_m$ is a 
gauge field on the D2-brane world-volume and couples to the R-R 1-form $C^{(1)}_m$, 
NS-NS 2-form $B_{mn}$ through the following couplings in the D2-brane 
effective action $S_{D2}$,
\begin{eqnarray}
S_{D2} \sim \int \! d^3 x \ 
\left[
B_{0i} F'^{0i} + F'_{12} C^{(1)}_0 + \cdots
\right].
\end{eqnarray}
Therefore, the electric field is a source of F-strings while the magnetic 
field is a source of D0-branes. 
From this observation, the above solutions we have obtained can be interpreted as 
an M2-M2 bound state which is reduced to D2-F1 bound state in the type 
IIA limit. Under the BPS conditions (\ref{BPSeq}), the supersymmetry transformation of 
the fermion becomes
\begin{eqnarray}
0 = \delta \psi_A = - \gamma^1 \omega_{AB} D_1 Y^B - \gamma^2 \omega_{AB} 
D_2 Y^B 
= 
- (\gamma^1 \pm i \gamma^2) \omega_{AB} D_1 Y^B,
\end{eqnarray}
namely, 
\begin{eqnarray}
\gamma^1 \gamma^2 \omega_{AB} = \pm i \omega_{AB}.
\end{eqnarray}
Therefore the BPS conditions (\ref{BPSeq}) preserve $1/2$ supersymmetry 
among $\mathcal{N} = 6$ supersymmetry.
We will see that this is an exact solution even for the full non-linear 
case. 

\section{Non-linear extension of the abelian ABJM model}
\subsection{Non-linear action}
Since the effective theory of an M2-brane without gauge fields is 
described by the Nambu-Goto action \cite{BeSeTo}, the scalar field part of the 
non-linear ABJM action is given by 
\begin{eqnarray}
S_{\mathrm{NG}} = \int \! d^3 x \ \mathcal{L}_{\mathrm{NG}}
= - T_{M2} \int \! d^3 x \ 
\sqrt{
- \det 
\left(
\eta_{m n} + T^{-1}_{M2} D_{(m} Y^A D_{n)} Y^{\dagger}_A 
\right)
}, 
\label{M2DBI}
\end{eqnarray}
where we have gauge covariantized the Nambu-Goto action by the gauge symmetry $U(1) 
\times U(1)$ and the parentheses in the determinant stands for the 
symmetrization of indices. 
At leading order in $T_{M2}^{-1}$, the action reduces to the scalar kinetic 
term in the abelian ABJM model (\ref{ABJM_action}). 
Let us analyze how this action is related to 
the D2-brane via the higgs mechanism.
The determinant factor $X \equiv \det (\delta_m {}^n + T_{M2}^{-1} D_m Y^A D^n Y^{\dagger}_A + D^n Y^A 
D_m Y^{\dagger}_A) $ in the squire root can be easily evaluated and its 
explicit form is given in Appendix.
Once a scalar field develops its VEV and is shifted around the vacuum, 
$Y^A \to v \delta^A {}_4 + Y^A$, and by taking the rescaling $B'_m \to 
B'_m/v$ and limits $k, v \to \infty$ (we call this process D2-reduction procedure), 
the equation of motion for the auxiliary field $B'_m$ is 
\begin{eqnarray}
0 = \frac{\partial \mathcal{L}'_{NG}}{\partial B'_m} = - \frac{T_{M2}}{2 \sqrt{X'}} \frac{\partial X'}{\partial B'_m}.
\end{eqnarray}
where $\mathcal{L}'_{\mathrm{NG}}$ is the Lagrangian after the 
D2-reduction procedure and we have defined the following quantity
\begin{eqnarray}
X' \equiv - \det \left[
\eta_{mn} + T_{M2}^{-1} D'_{(m} Y^A D'_{n)} Y^{\dagger}_A 
\right].
\end{eqnarray}
Here, $D'_m$ is the gauge covariant derivative of the scalar field after the introduction of 
the VEV. The explicit form of $\partial X'/ \partial B'_m$ is 
found in Appendix.
We find that a solution to the equation of motion is given by 
\begin{eqnarray}
B'_m = - \frac{\sqrt{2}}{4} \partial_m X^8,
\label{sol1}
\end{eqnarray}
which is the same one with the case in the linear order. 
Once this solution is substituted into the action and one rescales the scalar 
fields as $X^I \to g_{D2}^{-1} X^I$, we have 
\begin{eqnarray}
\mathcal{L}'_{\mathrm{NG}} &=& - T_{D2} \sqrt{\det (1+ T^{-1}_{D2} g^{-2}_{D2} \sum_{I = 1}^7 \partial_m X^I \partial^m X^I)} 
\label{D2NG}
\end{eqnarray}
Note that we have used the fact that the tension $T_{D2}$ of a D2-brane 
is obtained from $T_{M2}$ via the relation $T_{M2} = \frac{1}{(2\pi)^2} 
M_{11}^3 = \frac{1}{(2\pi)^2} g^{-1}_s \alpha^{\prime - \frac{3}{2}} = 
T_{D2}$, where the eleven-dimensional Planck mass $M_{11}$ is evaluated 
 by the ten-dimensional quantities.
The Lagrangian (\ref{D2NG}) is nothing but the scalar part of the DBI Lagrangian for 
the single D2-brane.

Since the gauge fields do not 
propagate in the ABJM model, it must not have kinetic terms in the non-linear theory 
and have topological terms only. 
Therefore, the full non-linear extension of the abelian ABJM action is given by 
\begin{eqnarray}
S_{M2} = \int \! d^3 x \ 
\left[
\frac{k}{4\pi} \epsilon^{m n p} (A_{m} 
\partial_{n} A_{p} - \hat{A}_{m} \partial_{n} \hat{A}_{p}) 
- T_{M2} 
\sqrt{
- \det 
\left(
\eta_{m n} + T^{-1}_{M2} D_{(m} Y^A D_{n)} Y^{\dagger}_A 
\right)
}
\right]
\label{ABJMDBI}
\end{eqnarray}
where we have considered only the bosonic part of the action. The 
fermionic part is given by supersymmetrizing this action.
A similar action was discussed in \cite{AlMu} but here we will study the 
detail structure of this action.
First of all, this action is reduced to the $U(1) \times U(1)$ ABJM 
model at the leading order in $T_{M2}^{-1}$ and to the Nambu-Goto action 
when all the gauge fields are dropped. The action is invariant under the 
$SU(4)_R$ R-symmetry, $U(1) \times U(1)$ gauge symmetry 
(\ref{gauge_transf}) 
and the orbifold projection $Y^A \to e^{i \frac{2\pi}{k}} Y^A$.
Note that the action proposed here does not contain 
higher derivative parts of the gauge fields that correspond to the derivative 
corrections of $F_{mn}$ in the D-brane effective action.
Actually, these terms will be absent when we will reduce the action to that
 of the D2-brane by the D2-reduction procedure. 
As in the case of the ordinary DBI action, we introduce a
 Lagrange multiplier $p$ and rewrite the action (\ref{ABJMDBI}) as follows,
\begin{eqnarray}
S_{M2} 
&=& \int \! d^3 x \ 
\left[
\frac{k}{4\pi} \epsilon^{m n p} (A_{m} 
\partial_{n} A_{p} - \hat{A}_{m} \partial_{n} \hat{A}_{p}) 
\right.
\nonumber \\
& & \left.
+ \frac{T_{M2}^2}{2p} \det \eta - \frac{p}{2}
\det 
\left(
\delta_m {}^n 
+ 
T_{M2}^{-1} D_m Y^A D^n Y^{\dagger}_A 
+ T_{M2}^{-1} D^n Y^A D_m Y^{\dagger}_A 
\right)
\right].
\label{ABJMDBI2}
\end{eqnarray}
If $p$ is integrated out, the original action (\ref{ABJMDBI}) is recovered.
With the above results in mind, 
let us consider the D2 reduction of the action (\ref{ABJMDBI2}). 
After the D2-reduction procedure, the action becomes
\begin{eqnarray}
S_{D2} &=& \int \! d^3 x \ 
\left[
\frac{k}{2\pi v} \epsilon^{mnp} 
F'_{mn} B'_p 
+ \frac{T_{D2}^2}{2p} \det \eta - \frac{p}{2} X' (B')
\right],
\label{D2aux}
\end{eqnarray}
where $X'$ is the determinant factor as we have defined before which is a 
function of $B'_m$. 
The equation of motion for the auxiliary field is
\begin{eqnarray}
\frac{k}{2\pi v} \epsilon_{mpq} F^{\prime pq} - \frac{1}{2} p \frac{\partial 
X'}{\partial B'_m} = 0.
\label{eom}
\end{eqnarray}
Since the solution in the absence of the 
Chern-Simons term is given by the equation~(\ref{sol1}) , 
we assume that the solution to the equation~(\ref{eom}) is given by the following form,
\begin{eqnarray}
B'_m = - \frac{\sqrt{2}}{4} \partial_m X^8 + \frac{1}{2\sqrt{2}} b_m.
\label{ansatz}
\end{eqnarray}
We are going to determine the function $b_m$. For an arbitrary function $f(B'_m)$, we 
have the following relation.
\begin{eqnarray}
\frac{\partial}{\partial B'_m} f (B'_m) \large|_{B'= - \frac{\sqrt{2}}{4} 
\partial X^8 + b/2\sqrt{2}} = 2 \sqrt{2} \frac{\partial}{\partial b_m} 
f (- \frac{\sqrt{2}}{4} \partial_m X^8 + \frac{b_m}{2\sqrt{2}}), 
\end{eqnarray}
Using this relation, the equation for the $b$ field is given by
\begin{eqnarray}
\frac{k}{2\pi v} \epsilon_{mpq} F^{\prime pq} - \sqrt{2} p 
\frac{\partial}{\partial b_m} X' (- \frac{\sqrt{2}}{4} \partial_m X^8 + 
\frac{b_m}{2\sqrt{2}}) = 0.
\label{b_equation}
\end{eqnarray}
Once we define $g_{mn} = \eta_{mn} + T^{-1}_{M2} \sum_{I=1}^7 \partial_m X^I 
\partial_n X^I$, we have 
\begin{eqnarray}
X' (- \frac{\sqrt{2}}{4} \partial_m X^8 + 
\frac{1}{2 \sqrt{2}} b_m) &=& - \det (\eta_{mn} + T^{-1}_{M2} \sum_{I=1}^7 \partial_m X^I 
\partial_n X^I + T_{D2}^{-1} b_m b_n) 
\nonumber \\
&=& - \det g \det (\delta_m {}^n + T_{D2}^{-1} b_m b_p g^{np}) 
\nonumber \\
&=& - \det g (1 + T_{D2}^{-1} b_m b_n g^{mn}).
\label{b_rel}
\end{eqnarray}
Therefore, 
\begin{eqnarray}
\frac{\partial}{\partial b_m} X' (- \frac{\sqrt{2}}{4} \partial_m X^8 + 
\frac{1}{2 \sqrt{2}} b_m) 
= - 2 T_{D2}^{-1} \det g \cdot g^{mn} b_n.
\end{eqnarray}
Then, the solution to the equation (\ref{b_equation}) is given by 
\begin{eqnarray}
b_m = - \frac{T_{D2}}{2\sqrt{2} p} \frac{1}{\det g} \frac{k}{2\pi v} g_{mr} 
\epsilon^{rpq} F'_{pq}.
\label{b_solution} 
\end{eqnarray}
At the end, we have a solution for $B'_m$, 
\begin{eqnarray}
B'_m = - \frac{\sqrt{2}}{4} \partial_m X^8 
- \frac{T_{D2}}{2\sqrt{2} p} \frac{1}{\det g} \frac{k}{2\pi v} g_{mr} 
\epsilon^{rpq} F'_{pq}.
\end{eqnarray}
If we substitute the solution 
$B'_m = - \frac{\sqrt{2}}{4} \partial_m X^8 + \frac{1}{2 \sqrt{2}} b_m$
 back into the action (\ref{D2aux}), we have 
\begin{eqnarray}
S_{D2} &=& \int \! d^3 x 
\left[
\frac{k}{2 \pi v} \epsilon^{mnp} F'_{mn} 
\left(
- \frac{\sqrt{2}}{4} \partial_p X^8 + \frac{1}{2 \sqrt{2}} b_p
\right) 
+ \frac{T_{D2}^2}{2p} \det \eta - \frac{p}{2} 
X' (
- \frac{\sqrt{2}}{4} \partial_m X^8 + \frac{1}{2 \sqrt{2}} b_m
) \right]
 \nonumber \\
&=& \int \! d^3 x \ 
\left[
 \frac{k}{4 \sqrt{2} \pi v} \epsilon^{mnp} F'_{mn} b_p + \frac{p}{2} 
\det g (1 + T_{D2}^{-1} b_m b_n g^{mn}) + \frac{T_{D2}^2}{2p} \det \eta
\right].
\end{eqnarray}
The term $\epsilon^{mnp} F'_{mn} \partial_p X^8$
becomes the total derivative by using the Bianchi identity for $F'_{mn}$. 
Rescaling $\frac{1}{g_{D2}} b_m = t_m$, we have 
\begin{eqnarray}
S_{D2} = \int \! d^3 x \ 
\left[
\frac{p}{2} 
\det g \left(1 + \lambda^2 g^4_{D2} t_m t_n g^{mn} \right) 
+ \frac{1}{2} \epsilon^{mnp} t_m F'_{np} - \frac{T^2_{D2}}{2p}
\right].
\end{eqnarray}
This is nothing but the dual D2 DBI action (\ref{D2DBI3}). 
Therefore the proposed action (\ref{ABJMDBI}) correctly reproduces the 
D2-brane action containing all $\alpha'$ corrections.

\subsection{BPS solutions}
Let us find BPS conditions in this non-linear action. 
When we consider effective theories of D-branes, it is important to bear 
in mind that BPS configurations in the lowest order in the derivative 
expansion is also an exact solution in the full non-linear order.
As a matter of fact, we will see that this is true even in our case.
We consider a configuration $Y^1 = Y \not=0, Y^{A\not=1} = 0$, then the 
determinant factor reduces to
\begin{eqnarray}
X = 1 + 2 T_{M2}^{-1} D_m Y D^m Y^{\dagger} 
+ T_{M2}^{-2} (D_m Y D^m Y^{\dagger})^2 - T_{M2}^{-2} D_m Y D^m Y D_n 
Y^{\dagger} D^n Y^{\dagger}.
\end{eqnarray}
From this expression, we have the following energy density
\begin{eqnarray}
\mathcal{E} = \frac{T_{M2}}{\sqrt{X}}
\left[
1 + 2 T_{M2}^{-1} D_i Y D_i Y^{\dagger} + T_{M2}^{-2} (D_i Y D_i 
Y^{\dagger})^2 - T_{M2}^{-2} D_i Y D_i Y D_j Y^{\dagger} D_j Y^{\dagger}
\right].
\label{NLenergy}
\end{eqnarray}
The determinant factor can be rewritten as 
\begin{eqnarray}
X &=& \left|
\frac{}{}
1 \pm i \epsilon^{ij} T_{M2}^{-1} D_i Y D_j Y^{\dagger}
\right|^2 + T_{M2}^{-1} 
\left|
D_i Y \pm i \epsilon_{ij} D_j Y
\right|^2 
\nonumber \\
& & \qquad \qquad 
- 2 T_{M2}^{-1} |D_0 Y|^2 
- T_{M2}^{-2} 
\left|
D_0 Y D_i Y^{\dagger} - D_0 Y^{\dagger} D_i Y
\right|^2.
\end{eqnarray}
while the expression in the bracket in the equation (\ref{NLenergy}) is rewritten as 
\begin{eqnarray}
\left|
1 \pm i \epsilon^{ij} T_{M2}^{-1} D_i Y D_j Y^{\dagger}
\right|^2 + T_{M2}^{-1} 
\left|
D_i Y \pm i \epsilon_{ij} D_j Y
\right|^2.
\end{eqnarray}
From these results, we find that the energy density is bounded as
\begin{eqnarray}
\mathcal{E} &\ge& 
T_{M2} 
\sqrt{
\left|
1 \pm i \epsilon^{ij} T_{M2}^{-1} D_i Y D_j Y^{\dagger}
\right|^2 + T_{M2}^{-1} 
\left|
D_i Y \pm i \epsilon_{ij} D_j Y
\right|^2 
}
\nonumber \\
&\ge& T_{M2} \left|
1 \pm i \epsilon^{ij} T_{M2}^{-1} D_i Y D_j Y^{\dagger}
\right|.
\end{eqnarray}
Here the equality holds only when the following conditions are satisfied,
\begin{eqnarray}
D_0 Y = 0, \qquad D_i Y \pm i \epsilon_{ij} D_j Y = 0.
\label{nlbps}
\end{eqnarray}
These are nothing but the BPS conditions found in the abelian ABJM model. 
If these conditions are satisfied, the energy is evaluated as 
\begin{eqnarray}
E = T_{M2} \int d^2 x 
\left(
1 + T_{M2}^{-1} |D_i Y|^2
\right)
= 
\pm i 
\epsilon_{ij} \int \! d^2 x \ 
\partial_i (Y^A D_j Y^{\dagger}_A),
\end{eqnarray}
where in the second equality, we have subtracted the mass of an M2-brane.
The surface term is the same one found in the abelian ABJM model and it 
is a generalization of the well-known M2-brane central 
charge \cite{GaGoTo, Lo}. 
The constraint coming from the $A_m$ gauge field equation of motion is 
rather complicated compared with the ABJM case (\ref{CS_constraint}).
The explicit form is found in Appendix.
However, if we consider the configuration $Y^1 = Y \not=0, Y^A = 0, 
(A\not=1)$ and use the BPS conditions, the Chern-Simons constraint (\ref{NLCS_constraints_app})
 becomes
\begin{eqnarray}
\frac{k}{4\pi} \epsilon^{mpq} F_{pq} - i (Y D^m Y^{\dagger} - Y^{\dagger} 
D^m Y) = 0.
\label{NLCS_constraints}
\end{eqnarray}
The same form of the condition holds for the gauge field $\hat{A}_m$ (just replace $F_{mn} \to \hat{F}_{mn}$).
These conditions precisely realize the Chern-Simons conditions obtained in the linear order. 
Therefore the BPS solutions in the abelian ABJM model are the exact 
solutions even in the non-linear order. This situation is quite similar to the 
D-brane case \cite{CaMa, CoMyTa}. All higher derivative effects in 
the non-linear action cancel for the BPS configurations.

\section{Couplings with supergravity backgrounds}
\subsection{Linear order}
The coupling of a single M2-brane with the 3-form background 
in the eleven-dimensional supergravity in flat space is given by \cite{BeSeTo}.
\begin{eqnarray}
S &=& T_{M2} \int \! d^3 x \ \epsilon^{mnp} P[C^{(3)}]_{mnp} 
\nonumber \\
 &=& T_{M2} \int \! d^3 x \ \epsilon^{mnp} 
\left[
C^{(3)}_{mnp} + 3 \partial_m X^I C^{(3)}_{Inp} 
+ 3 \partial_m X^I \partial_n X^J C^{(3)}_{IJp} 
+ \partial_m X^I \partial_n X^J 
\partial_p X^K C^{(3)}_{IJK}
\right],
\nonumber \\
\label{3-form}
\end{eqnarray}
where the real scalars $X^I \ (I=1, \cdots, 8)$ stand for the transverse directions to 
the M2-brane world-volume and the pull-back has been evaluated in the static gauge.
We generalize this term to the abelian ABJM model. 
For the ABJM model, the transverse direction is $\mathbf{Z}_k$-orbifolded and part of 
the components in the 3-form in the real basis $C^{(3)} = \frac{1}{3!} C^{(3)}_{MNP} d x^M \wedge d 
x^N \wedge d x^P, \ (M,N,N=0, \cdots 10)$ should be projected out. 
In the static gauge, there are four index structures\footnote{In the 
following, we omit the superscript ``$(3)$'' for the 3-form in eleven dimensions.} 
$C_{IJK}, C_{mIJ}, C_{mnI}, C_{mnp}$ where $I,J,K = 1, \cdots, 8$ are 
transverse directions and $m,n,p = 0,1,2$ are world-volume directions. 
The components which contain odd number of indices $I,J,K$ 
are projected out while $C_{mnp}$ is apparently invariant under the orbifold. 
For the index structure $C_{mIJ}$, focusing on the transverse directions, 
the 3-form can be rewritten by the complex coordinate basis $y^A = \frac{1}{\sqrt{2}} (x^A + i x^{A+4}), 
y^{\dagger}_A = \frac{1}{\sqrt{2}} (x^A - i x^{A+4}), \ A, B = 1, \cdots, 
4$. From the orbifold condition $y^A \to e^{\frac{2\pi i}{k}} y_A$, we 
find that the conditions of the 3-form in the $x$-basis components are 
given by
\begin{eqnarray}
C_{mAB} = C_{m(A+4)(B+4)}, \quad C_{m A(B+4)} = - C_{m (A+4)B}.
\label{projection}
\end{eqnarray}
Therefore the components that survive the orbifold projection are 
found to be 
\begin{eqnarray}
\mathcal{C}_{mA} {}^B &\equiv& 
\frac{1}{2} 
\left\{
C_{mAB} + i C_{m A (B+4)} - i C_{m (A+4) B} + C_{m (A+4) (B+4)}
\right\}, \\
\mathcal{C}_m {}^A {}_B 
&\equiv& 
\frac{1}{2} 
\left\{
C_{mAB} - i C_{m A (B+4)} + i C_{m (A+4) B} + C_{m (A+4) (B+4)}
\right\}, \\
(\mathcal{C}_{mA} {}^B)^{\dagger} &=& \mathcal{C}_m {}^A {}_B. 
\end{eqnarray}
We propose coupling terms of the 3-form with the abelian ABJM model 
as 
\begin{eqnarray}
S_{\mathrm{flux}} = 
\frac{T_{M2}}{3!} \int \! d^3 x \ \epsilon^{mnp} C_{mnp} + 
\frac{1}{2} \int \! d^3 x \ \epsilon^{mpq} 
\left[
\mathcal{C}_{mA} {}^B  D_p Y^A D_q Y^{\dagger}_B
+ \mathcal{C}_m {}^A {}_B D_p Y^{\dagger}_A D_q Y^B
\right],
\label{flux}
\end{eqnarray}
where $\mathcal{C}_{mA} {}^B, \mathcal{C}_m 
{}^A {}_B$ transform as $\bar{\mathbf{4}} \otimes \mathbf{4}, \mathbf{4} 
\otimes \bar{\mathbf{4}}$ under the $SU(4)_R$ R-symmetry. 
All the scalar fields $Y^A$ appearing in the ABJM model is rescaled in 
such a way that the overall factor $T_{M2}$ in (\ref{3-form}) is absent.
The action (\ref{flux}) is invariant under 
the world-volume $U(1) \times U(1)$ gauge transformation. 
This is the natural generalization of the ordinary 3-form coupling 
(\ref{3-form}). The structure of the background field and the gauge 
invariance is consistent and the pull-back is gauge covariantized which 
is similar to the Myers term in the D-brane cases \cite{My}. 
The first term in the equation (\ref{flux}) trivially reduces to the natural 
world-volume coupling of the R-R 3-form $C^{(3)}_{mnp}$ with the D2-brane, namely, the first term in (\ref{RR3}). 
Note that we are considering the constant potential for simplicity and 
it breaks the gauge invariance of the background field $\delta C^{(3)} = d 
\Lambda^{(2)}$ where $\Lambda^{(2)}$ is a 2-form gauge parameter.
If the background is not constant, it should be expanded 
around the M2-brane position $X_0$. For example, in the $x$-basis, the 
potential $C_{IJK}$ is expanded as 
\begin{eqnarray}
C_{IJK} (X) = C_{IJK} (X_0) + \partial_R C_{IJK} (X_0) X^R + \frac{1}{2!} 
\partial_S \partial_R C_{IJK} (X_0) X^R X^S + \cdots.
\end{eqnarray}
%
Let us see the D2 reduction of the action~(\ref{flux}). We start with the action 
\begin{eqnarray}
\tilde{S} = S_{\mathrm{ABJM}} + S_{\mathrm{flux}},
\label{flux_action}
\end{eqnarray}
where the first term is the action for the abelian ABJM model (\ref{ABJM_action}). 
After the D2-reduction procedure, we have the flux part 
$S'_{\mathrm{flux}}$ coming from $S_{\mathrm{flux}}$,
\begin{eqnarray}
S'_{\mathrm{flux}} &=& \frac{T_{D2}}{3!} \int \! d^3 x \ \epsilon^{mnp} 
C^{(3)}_{mnp} + 
\int \! d^3 x \ \frac{1}{2} \epsilon^{mpq} 
\left[ \frac{}{}
\mathcal{C}_{mA} {}^B \partial_p Y^A \partial_q Y^{\dagger}_B - 
\mathcal{C}_m {}^A {}_B \partial_p Y^B \partial_q Y^{\dagger}_A 
\right. 
\nonumber \\
& & \left. \frac{}{}
+ 2 i (\mathcal{C}_{mA} {}^4 - \mathcal{C}_m {}^4 {}_A ) B'_p \partial_q Y^A
+ 2 i (\mathcal{C}_{m4} {}^A - \mathcal{C}_m {}^A {}_4 ) B'_p \partial_q Y^{\dagger}_A
\right].
\end{eqnarray}
The equation of motion for the auxiliary field is 
solved by the following solution,
\begin{eqnarray}
B^{\prime p} &=& - \frac{\sqrt{2}}{4} \partial^p X^8 + \frac{1}{8} \frac{k}{2\pi v} 
\epsilon^{mnp} F'_{mn} 
\nonumber \\
& & -  \frac{\sqrt{2}}{8} \epsilon^{mpq} 
\left[
C_{mA4} \partial_q X^{A+4} + C_{m (A+4) 8} \partial_q X^{A+4} 
+ C_{m4(A+4)} \partial_q X^A + C_{mA8} \partial_q X^A
\right].
\nonumber \\
\end{eqnarray}
Substituting this solution back into the action (\ref{flux_action}), we find that all terms 
which contain $X^8$ cancel out and the resulting action $\tilde{S}_{D2}$ 
contains following coupling terms, 
\begin{eqnarray}
\tilde{S}_{D2} &\sim& 
\int \! d^3 x \ \left[
\frac{T_{D2}}{3!} \epsilon^{mnp} C^{(3)}_{mnp} + 
\frac{1}{2 g_{D2}^2} \epsilon^{mpq} \sum_{I,J=1}^7 C^{(3)}_{mIJ} \partial_p X^I 
\partial_q X^J - \frac{1}{g_{D2}^2} F^{\prime mn} \sum_{I=1}^7 
B_{mI} \partial_n X^I 
\right.
\nonumber \\
& & \left.
- \frac{1}{2 g^2_{D2}} \sum_{I,J=1}^7 B_{mI} \partial_p X^I B^m {}_{J} \partial^p X^J
+ \frac{1}{2 g^2_{D2}} \sum_{I,J=I}^7 B_{mI} \partial_n X^I B^n {}_{J} \partial^m X^J
\right],
\end{eqnarray}
where we have used the orbifold projection condition (\ref{projection}), 
defined the NS-NS 2-form as $C_{mI8} \equiv B_{mI}$ and rescaled $X^I \to 
g_{D2}^{-1} X^I$. This result precisely reproduces part of the couplings in the 
D2-brane action (\ref{RR3}), (\ref{NSNS2}).
Note that other parts which are absent in the above 
expression, for example terms that contain $B_{mn}$ in 
(\ref{NSNS2}), have been already projected out in the eleven dimensions and 
never appear in ten dimensions. 

\subsection{Non-linear extension}
Let us generalize the previous result to the non-linear case. 
We propose the following action,
\begin{eqnarray}
\hat{S}_{M2} &=& \int \! d^3 x \ \left[
\frac{k}{4\pi} \epsilon^{mnp} 
(A_m \partial_n A_p - \hat{A}_m \partial_n \hat{A}_p)
\right.  \nonumber \\
& & + \frac{T_{M2}^2}{2p} \det \eta - \frac{p}{2} \det 
\left(
\delta_m {}^n + T_{M2}^{-1} D_m Y^A D^n Y^{\dagger}_A + T_{M2}^{-1} D^n 
Y^A D_m Y^{\dagger}_A
\right) \nonumber \\
& & \left. 
+ \frac{T_{M2}}{3!} \epsilon^{mnp} C_{mnp}
+ \frac{1}{2} \epsilon^{mpq} 
\left[
\mathcal{C}_{mA} {}^B D_p Y^A D_q Y^{\dagger}_B + \mathcal{C}_m {}^A 
{}_B D_p Y^{\dagger}_A D_q Y^B
\right]
\right],
\end{eqnarray}
where $p$ is the Lagrange multiplier as we have introduced before.
After the D2-reduction procedure, we have the action $\hat{S}_{D2}$ 
which is given by 
\begin{eqnarray}
\hat{S}_{D2} &=& \int \! d^3 x \ 
\left[
\frac{k}{2\pi v} \epsilon^{mnp} 
F'_{mn} B'_p 
+ \frac{T_{D2}^2}{2p} \det \eta - \frac{p}{2} X' + 
\frac{1}{2} \epsilon^{mpq} 
\left[
\mathcal{C}_{mA} {}^B D_p Y^A D_q Y^{\dagger}_B + \mathcal{C}_m {}^A 
{}_B D_p Y^{\dagger}_A D_q Y^B
\right] 
\right.
\nonumber \\
& & 
\left. \frac{}{}
+ \frac{T_{D2}}{3!} \epsilon^{mnp} C^{(3)}_{mnp}
+ i \epsilon^{mpq} 
\left(
\mathcal{C}_{mA} {}^4 - \mathcal{C}_m {}^4 {}_A 
\right)
B'_p \partial_q Y^A 
+ i \epsilon^{mpq} 
\left(
\mathcal{C}_{m4} {}^A - \mathcal{C}_m {}^A {}_4
\right) B'_p \partial_q Y^{\dagger}_A
\right].
\label{reduced_D2}
\end{eqnarray}
The equation of motion for the auxiliary field is 
\begin{eqnarray}
\frac{\partial \hat{\mathcal{L}}_{D2}}{\partial B'_p} &=& \frac{k}{2\pi v} 
\epsilon^{mnp} F'_{mn} 
- \frac{p}{2} \frac{\partial X'}{\partial B'_p} 
+ i \epsilon^{mpq} 
\left(
\mathcal{C}_{mA} {}^4 - \mathcal{C}_m {}^4 {}_A 
\right)
\partial_q Y^A 
+ i \epsilon^{mpq} 
\left(
\mathcal{C}_{m4} {}^A - \mathcal{C}_m {}^A {}_4
\right) \partial_q Y^{\dagger}_A,
\nonumber \\
\end{eqnarray}
where $\hat{\mathcal{L}}_{D2}$ is the Lagrangian corresponding to the 
action (\ref{reduced_D2}). 
As in the case of section 4, consider the ansatz (\ref{ansatz}) and using 
relations (\ref{b_rel}) and 
\begin{eqnarray}
\left(
\mathcal{C}_{mA} {}^4 - \mathcal{C}_m {}^4 {}_A 
\right)
\partial_q Y^A 
+
\left(
\mathcal{C}_{m4} {}^A - \mathcal{C}_m {}^A {}_4
\right) \partial_q Y^{\dagger}_A
= 2 \sqrt{2} i \sum_{I=1}^7 C_{mI8} \partial_q X^I,
\end{eqnarray}
the solution for the $b_m$ equation is found to be 
\begin{eqnarray}
b_m = - \frac{T_{D2}}{p \det g} g_{mp} \epsilon^{prs} 
\left[
\frac{1}{2 g_{D2}} F'_{rs} + \sum_{I=1}^7 C_{rI8} \partial_s X^I
\right].
\end{eqnarray}
This is just the solution (\ref{b_solution}) but with the replacement 
$\frac{1}{2 g_{D2}} F'_{mn} \to \frac{1}{2 g_{D2}} F'_{mn} + \sum_{I=1}^7 C_{mI8} \partial_n X^I$.
Substituting the ansatz into the action and using the orbifold condition, again we find that all the $X^8$ 
dependence cancel out and the action becomes 
\begin{eqnarray}
\hat{S}_{D2} &=& \int \! d^3 x \ 
\left[
\epsilon^{mnp} 
\left[
\frac{1}{2 g_{D2}} F'_{mn} + C_{mI8} \partial_n X^I
\right] b_p + \frac{T_{D2}^2}{2p} \det \eta + \frac{p}{2} \det g (1 + 
T_{D2}^{-1} b_m b_n g^{mn})
\right.
\nonumber \\
& & \left. 
+ \frac{T_{D2}}{3!} \epsilon^{mnp} C^{(3)}_{mnp}
+ \frac{1}{2} \epsilon^{mpq} 
\sum_{I,J=1}^7 C^{(3)}_{mIJ} \partial_p X^I \partial_q X^J
\right].
\label{D2DBI_d}
\end{eqnarray}
The first three terms are just the dual form of the D2 DBI action 
(\ref{D2DBI3}) and after integrating out $p$ and $b_m$, we find 
\begin{eqnarray}
\hat{S}_{D2} &=& - T_{D2} \int \! d^3 x \ 
\sqrt{
- \det 
\left(
g_{mn} + (B_{mI} \partial_n X^I - B_{nI} \partial_n X^I) + \lambda F'_{mn}
\right)
}
\nonumber \\
& & + \int \! d^3 x \ 
\left[
\frac{T_{D2}}{3!} \epsilon^{mnp} C^{(3)}_{mnp}
+ \frac{1}{2 g^2_{D2}} \epsilon^{mpq} \sum_{I,J=1}^7 C^{(3)}_{mIJ} \partial_p X^I 
\partial_q X^J
\right],
\nonumber \\
\end{eqnarray}
where we have rescaled $X^I \to g_{D2}^{-1} X^I$ and 
used the fact that $C_{mI8} = B_{mI}$ couples to anti symmetric tensors
in the determinant. This result has correct structure of the R-R 3-form 
and NS-NS 2-form couplings in the DBI action. 
It is worthwhile to emphasize that the correct (including numerical coefficients) structure 
of couplings for both R-R 3-form and NS-NS 2-form are generated by the 
higgs mechanism from the coupling of the 3-form in eleven dimensions.

\section{Conclusions and discussions}
In this paper, we investigated higher derivative corrections 
to the abelian ABJM model and its couplings with the 3-form background in 
eleven-dimensional supergravity. The orbifold projection singles out 
the gauge invariant combination of the couplings which is desired result 
from the viewpoint of the world-volume theory.
We showed that the novel higgs mechanism does work even in 
the presence of the higher derivative corrections and the supergravity 
coupling suggesting the ubiquity of the higgs mechanism in M2-brane 
effective theories. The 
equations of motion for the auxiliary fields are solved explicitly and the 
correct structure of the D2-brane effective action is obtained. 

We also studied BPS configurations that keep half of the $\mathcal{N} = 6$ 
supersymmetry. The solutions can be interpreted as an M2-brane 
intersecting with the original M2-brane. We showed that the BPS solution 
in the abelian ABJM model is also an exact solution in the non-linear 
theory. Similar to the D-brane case, all the higher derivative 
corrections cancel out at least for the BPS configurations. 
However, when one considers non-BPS configurations, these higher 
derivative corrections play a significant role to study the dynamics of 
M2-branes \cite{FuIwKoSa}. 

The results of the abelian case presented in this paper provide valuable 
intuition for the construction of the non-linear extension of 
non-abelian theories. 
A typical example is $\mathcal{N} = 6$ ABJM model with $U(N) \times 
U(N)$  gauge group. However, we find that 
the natural non-abelian extension of the abelian action (\ref{ABJMDBI}) 
fails due to the non-availability of the higgs mechanism -- $X^8$ does 
not cancel out in the naive D2-reduction procedure. 
Presumably, some symmetrization of the gauge trace need to be introduced.
There are several criteria for constructing a non-linear action for 
non-abelian gauge groups. First, at leading 
order in $T_{M2}^{-1}$, the action should reduces to the $U(N) \times U(N)$ 
ABJM model and also settle down to the action (\ref{ABJMDBI}) in the 
abelian limit. Second, assuming that the higgs mechanism works even in the non-abelian case, 
the action must reduce to the non-abelian DBI action proposed in 
\cite{My, Ts} which is correct at least up to order $\mathcal{O} (F^6)$ 
and naturally incorporates the symmetric trace structure.
Finally, we expect that BPS solutions which have been found in the $U(N) \times U(N)$ 
ABJM model are solutions even in the non-linear order. These 
conditions severely constrain the possible form of the action. 
Note that in the ABJM model, the symmetric trace structure of the fields 
is rather unclear since the fields $Y^A$ (and its supersymmetric 
counterpart) are all in the bi-fundamental representation of the gauge 
group. Therefore it would be better to study the structure of the action 
by order by order in $T_{M2}^{-1}$. 

On the other hand, we can consider the 3-form coupling with the 
non-abelian ABJM model.
A natural non-abelian generalization of the coupling 
(\ref{flux}) would be given by
\begin{eqnarray}
S_{\mathrm{flux}} = 
\frac{1}{2} \int \! d^3 x \ \epsilon^{mpq} 
\mathrm{Tr}
\left[
\mathcal{C}_{mA} {}^B  D_p Y^A D_q Y^{\dagger}_B
+ \mathcal{C}_m {}^A {}_B D_p Y^{\dagger}_A D_q Y^B
\right],
\label{NA_flux}
\end{eqnarray}
where we have dropped the $C_{mnp}$ term and 
$Y^A$ are bi-fundamental representation of the gauge group, $\mathcal{C}_{mA} {}^B, \mathcal{C}_m {}^A 
{}_B$ are constants for simplicity. As in the case of the abelian gauge 
group, the orbifold projection singles out the gauge invariant 
combination of the background. 
It is easy to find that via the 
D2-reduction procedure, the coupling (\ref{NA_flux}) reproduces part of 
the NS-NS 2-form and R-R 3-form couplings in multiple D2-brane effective 
action. Especially, the gauge covariantized pull-back 
structure \cite{My} is reproduced.
Another interesting issue is to find the higher order corrections of the 
3-form (and its 6-form dual) couplings to the M2-brane\footnote{See 
\cite{LaRi, KiKwNaTo} for the ralated discussions in the BLG model.} which, 
after the reduction to the 
D2-brane effective theory, should be related to the deformed gauge 
theories on the D3-branes in the presence of R-R backgrounds via 
T-dualities \cite{deformed}. 
We will come back to this issue in the future works \cite{Sa}.

\subsection*{Acknowledgements}
S.~S would like to thank S.~Terashima for notifying the solution in 
the abelian ABJM model, and especially for M.~Arai for the collaboration in the  
early stage of this project. The work of S.~S. is supported by the Japan Society for
the Promotion of Science (JSPS) Research Fellowship.

\begin{appendix}
\section{Useful formulae}
The $3 \times 3$ determinant factor in the square root of the action (\ref{M2DBI}) 
is evaluated as  
\begin{eqnarray}
X &\equiv& \det (\delta_m {}^n + T_{M2}^{-1} D_m Y^A D^n Y^{\dagger}_A + D^n Y^A 
D_m Y^{\dagger}_A) 
\nonumber \\
& & = 1 + 2 T^{-1}_{M2} D_m Y^A D^m Y^{\dagger}_A 
+ 2 T_{M2}^{-2} (D_m Y^A D^m Y^{\dagger}_A)^2 
- \frac{1}{2} T^{-2}_{M2} D_{(m} Y^A D_{n)} Y^{\dagger}_A 
D^{(m} Y^B D^{n)} Y^{\dagger}_B 
\nonumber \\
& & + \frac{4}{3} T^{-3}_{M2} (D_m Y^A D^m Y^{\dagger}_A)^3 
+ \frac{2}{3} T^{-3}_{M2} (D_m Y^A D^p Y^{\dagger}_A) (D_n Y^B D^m 
Y^{\dagger}_B) (D_p Y^C D^n Y^{\dagger}_C) 
\nonumber \\
& & - T_{M2}^{-3} (D_m Y^A D^m Y^{\dagger}_A) D_{(p} Y^B D_{q)} 
Y^{\dagger}_B D^{(p} Y^C D^{q)} Y^{\dagger}_C 
\nonumber \\
& & + 2 T_{M2}^{-3} (D_m Y^A D^p Y^{\dagger}_A) (D^m Y^B D_n Y^{\dagger}_B) 
(D_p Y^C D^n Y^{\dagger}_C).
\end{eqnarray}
Once $Y^4$ develops a VEV $v$ and taking the limits $v,k \to \infty$ 
after the rescaling the auxiliary field $B'_m \to B'_m/v$, we rewrite 
$X$ by $X'$ to distinguish the original determinant with that contains the VEV.
The derivative of $X'$ with respect to $B'_m$ is explicitly given by 
\begin{eqnarray}
\frac{\partial X'}{\partial B'_m} &=& 
2 T_{M2}^{-1} \left( 2 \sqrt{2} \partial_m X^8 + 8 B'_m \right)
\nonumber \\
& & + 4 T_{M2}^{-2} (D'_n Y^A D^{\prime n} Y^{\dagger}_A) 
\left( 2 \sqrt{2} \partial_m X^8 + 8 B'_m \right)
\nonumber \\
& & 
- 2 T_{M2}^{-2} (D^{\prime (p} Y^A D^{\prime m)} Y^{\dagger}_A) 
\left( 2 \sqrt{2} \partial_p X^8 + 8 B'_p \right) 
\nonumber \\
& & + 4 T_{M2}^{-3} (D'_n Y^A D^{\prime n} Y^{\dagger}_A)^2 
\left( 2 \sqrt{2} \partial_m X^8 + 8 B'_m \right)
\nonumber \\
& & + 2 T_{M2}^{-3} (D^{\prime p} Y^B D'_n Y^{\dagger}_B) (D^{\prime n} 
Y^C D^{\prime m} Y^{\dagger}_C) \left( 2 \sqrt{2} \partial_p X^8 + 8 B'_p \right)
\nonumber \\
& & + 2 T_{M2}^{-3} (D^{\prime }_n Y^B D^{\prime p} Y^{\dagger}_B) 
(D^{\prime m} Y^C D^{\prime n} Y^{\dagger}_C) \left( 2 \sqrt{2} \partial_p X^8 + 8 B'_p \right)
\nonumber \\
& & + 2 T_{M2}^{-3} (D^{\prime m} Y^B D^{\prime n} Y^{\dagger}_B) 
(D^{\prime p} Y^C D'_n Y^{\dagger}_C) \left( 2 \sqrt{2} \partial_p X^8 + 8 B'_p \right)
\nonumber \\
& & + 2 T_{M2}^{-3} (D'_n Y^B D^{\prime p} D^{\dagger}_B) (D^{\prime n} 
Y^C D^{\prime m} Y^{\dagger}_C) \left( 2 \sqrt{2} \partial_p X^8 + 8 B'_p \right)
\nonumber \\
& & - T^{-3}_{M2} D'_{(n} Y^B D'_{p)} Y^{\dagger}_B D^{\prime (n} Y^C 
D^{\prime p)} Y^{\dagger}_C \left( 2 \sqrt{2} \partial_m X^8 + 8 B'_m \right)
\nonumber \\
& & - 4 T^{-3}_{M2} D'_n Y^A D^{\prime n} Y^{\dagger}_A D^{\prime (p} 
Y^C D^{\prime m)} Y^{\dagger}_C \left( 2 \sqrt{2} \partial_p X^8 + 8 B'_p \right).
\end{eqnarray}
From this expression, it is clear that (\ref{sol1}) is a solution of the 
 equation of motion for the auxiliary field $B'_m$.

Let us evaluate the equation of motion of $A_m$ for the action 
(\ref{ABJMDBI}). A straightforward calculation leads to the following result,
\begin{eqnarray}
0 &=& \frac{k}{4\pi} \epsilon^{mpq} F_{pq} - \frac{1}{\sqrt{X}}
\left\{ \frac{}{}
i (Y^A D^m Y^{\dagger}_A - Y^{\dagger}_A D^m Y^A) 
+ 2 i T_{M2}^{-1} 
(Y^A D^m Y^{\dagger}_A - Y^{\dagger}_A D^m Y^A) 
D_n Y^B D^n Y^{\dagger}_B
\right. \nonumber \\
& & - i T_{M2}^{-1} 
(Y^A D_p Y^{\dagger}_A - Y^{\dagger}_A D_p Y^A) 
D^{(m} Y^B D^{p)} Y^{\dagger}_B 
+ 2 i T_{M2}^{-2} (Y^A D^m Y^{\dagger}_A - Y^{\dagger}_A D^m Y^A) 
(D_n Y^B D^n Y^{\dagger}_B)^2 
\nonumber \\
& & 
+ i T_{M2}^{-2} Y^A D^p Y^{\dagger}_A (D_r Y^B D^m Y^{\dagger}_B) (D_p 
Y^C D^r Y^{\dagger}_C) 
- i T_{M2}^{-2} Y^{\dagger}_A D^p Y^A (D_r Y^B D_p Y^{\dagger}_B) (D^m 
Y^C D^r Y^{\dagger}_C)
\nonumber \\
& & - \frac{i}{2} T_{M2}^{-2} (Y^A D^m Y^{\dagger}_A - Y^{\dagger}_A D^m Y^A) 
D_{(p} Y^B D_{q)} Y^{\dagger}_B D^{(p} Y^C D^{q)} Y^{\dagger}_C 
\nonumber \\
& & 
- 2 i T_{M2}^{-2} (Y^A D_p Y^{\dagger}_A - Y^{\dagger}_A D_p Y^A) 
D_{n} Y^B D^n Y^{\dagger}_B D^{(m} Y^C D^{p)} Y^{\dagger}_C
\nonumber \\
& & + i T_{M2}^{-2} (Y^A D_p Y^{\dagger}_A - Y^{\dagger}_A D_p Y^A) 
(D^m Y^B D_r Y^{\dagger}_B) (D^p Y^C D^r Y^{\dagger}_C)
\nonumber \\
& & 
+ i T_{M2}^{-2} (Y^A D_p Y^{\dagger}_A - Y^{\dagger}_A D_p Y^A) 
(D_r Y^B D^m Y^{\dagger}_B) (D^r Y^C D^p Y^{\dagger}_C)
\nonumber \\
& & \left. \frac{}{}
+ i T_{M2}^{-2} (D^m Y^A D^p Y^{\dagger}_A) Y^B D_r Y^{\dagger}_B (D_p 
Y^C D^r Y^{\dagger}_C) 
- i T_{M2}^{-2} (D_r Y^A D^p Y^{\dagger}_A) Y^{\dagger}_B D^r Y^B 
(D_p Y^C D^m Y^{\dagger}_C)
\right\}
\nonumber \\
\label{NLCS_constraints_app}
\end{eqnarray}
The same equation holds even for $\hat{A}_m$ if one replaces $F_{mn}$ by 
$\hat{F}_{mn}$ in the above expression. 
If we assume the configuration $Y^1 = Y \not= 0, Y^A = 0, \ (A=2,3,4)$ 
and using the BPS conditions (\ref{nlbps}), the equation~(\ref{NLCS_constraints_app}) reduces 
to the equation~(\ref{NLCS_constraints}).

\end{appendix}

\end{document}